\documentclass{aastex}
\usepackage{emulateapj5}

\begin{document}

\title{THE LOW VELOCITY WIND FROM THE CIRCUMSTELLAR MATTER AROUND THE B9V STAR $\sigma$ HERCULIS}
\author{C. H. Chen \& M. Jura}
\affil{Department of Physics and Astronomy, University of California,
       Los Angeles, CA 90095-1562; cchen@astro.ucla.edu; 
       jura@clotho.astro.ucla.edu}

\begin{abstract}
We have obtained \emph{FUSE} spectra of $\sigma$ Her, a nearby binary system, 
with a main sequence primary, that has a Vega-like infrared excess. We 
observe absorption in the excited fine structure lines \ion{C}{2}$^{*}$ at 
1037 \AA, \ion{N}{2}$^{*}$ at 1085 \AA, and \ion{N}{2}$^{**}$ at 1086 \AA \ 
that are blueshifted by as much as $\sim$30 km/sec with respect to the star. 
Since these features are considerably narrower than the stellar lines and 
broader than interstellar features, the \ion{C}{2} and \ion{N}{2} are 
circumstellar. We suggest that there is a radiatively driven wind, arising 
from the circumstellar matter, rather than accretion as occurs around $\beta$ 
Pic, because of $\sigma$ Her's high luminosity. Assuming that the gas is 
liberated by collisions between parent bodies at 20 AU, the approximate 
distance at which blackbody grains are in radiative equilibrium with the star 
and at which 3-body orbits become unstable, we infer 
dM/dt $\sim$ 6 $\times$ $10^{-12}$ M$_{\sun}$ yr$^{-1}$. This wind depletes 
the minimum mass of parent bodies in less than the estimated age of the 
system. 
\end{abstract}


\section{INTRODUCTION}
Planets are believed to form within circumstellar disks around young main 
sequence stars (with ages $\leq$100 Myr) composed of dust and gas (Beckwith \&
Sargent 1996). Dusty circumstellar disks, with radii comparable to the 
distance between the Sun and the Kuiper Belt, have been imaged around several 
young, nearby main sequence stars (Schneider et al. 1999, Weinberger et al. 
1999, Holland et al. 1998). However, few studies have focused on the 
properties of gas and dust at a few AU from main sequence stars where the 
temperature of parent bodies may reach $\sim$300 K (Chen \& Jura 2001, Heap et
al. 2000, Low et al. 1999). At these distances, ice sublimates and large 
solids may evolve into terrestrial planets. We have carried out a far 
ultraviolet study of gas around a main sequence star, which possess a possible
10 $\mu$m excess attributed to dust with a grain temperature ($T_{gr}$ $\sim$ 
300 K), to learn about the physical processes in the regions where terrestrial
planets may have formed or may be forming.

$\sigma$ Herculis is a binary system, at a distance 93 pc away from the Sun 
(see Table 1), with a B9V primary and a companion at a projected separation of 
0.07$\arcsec$ (Hartkopf et al. 1997). The spectral type of the companion has 
not been well determined. Astrometric estimates for the masses, using orbital 
parameters determined from speckle interferometry and relative brightness 
measurements from \emph{Hipparcos}, yield M$_{1}$ = 3.0 $\pm$ 0.7 M$_{\sun}$, 
M$_{2}$ = 1.5 $\pm$ 0.5 M$_{\sun}$, and M$_{tot}$ = 4.5 $\pm$ 0.8 M$_{\sun}$ 
(Martin et al. 1998). The $V - R$ color and the absolute $V$-band magnitude of 
the secondary are consistent with a classification of early-type A ($\Delta m$ 
= 2.5 $\pm$ 0.1 mag for 5000 \AA \ $<$ $\lambda$ $<$ 8500 \AA; Hummel et al. 
2002). The age of $\sigma$ Her has been estimated to be $\sim$200 Myr based 
upon its uvby$\beta$ photometry (Grosb{\o}l 1978). If we apply a correction for
the high rotational velocity of this star to the Str\"{o}mgren photometry 
(Figueras \& Blasi 1998), we find an estimated age of $\sim$140 $\pm$ 100 Myr 
for $\sigma$ Her.

The $\sigma$ Her binary system possess a mid infrared excess indicative of 
the presence of dust ($L_{IR}/L_{*}$ = 6.6$\times$10$^{-5}$; Fajardo-Acosta, 
Telesco, \& Knacke 1998). The dust appears to be divided into two populations. 
The colder grains were first discovered based upon measurements of a strong 
\emph{IRAS} 60 $\mu$m excess (Sadakane \& Nishida 1986; Cot\'{e} 1987)
which is unresolved with \emph{ISO} (Fajardo-Acosta, Stencel, \& Backman 1997).
Recent ground based photometry of $\sigma$ Her suggests that this star may 
also possess 10 $\mu$m and 20 $\mu$m excesses (Fajardo-Acosta et al. 1998). 
The spectral energy distribution of this warmer dust population is marginally 
fit by blackbody grains with a temperature, $T_{gr} \sim$ 300 $\pm$ 100 K,
significantly warmer than $T_{gr}$ $\sim$ 100 K grains observed with 
\emph{IRAS} at 60 $\mu$m. If the particles are blackbodies in radiative 
equilibrium with the binary, then the dust is located at a distance of between 
7 AU and 30 AU, significantly closer than the 120 AU distance inferred for the 
cooler population. 

While Vega-like systems, such as $\sigma$ Her, possess dust, the gas:dust ratio
is not well known. Recent ultraviolet searches for molecular hydrogen, using 
\emph{FUSE}, have failed to detect any molecular hydrogen around the dusty main
sequence stars $\beta$ Pic (Lecavelier des Etangs et al. 2001) and 51 Oph 
(Roberge et al 2002). However, spectra of these systems possess time-variable, 
redshifted atomic gas features which are believed to be generated by the 
evaporation of infalling comets (Vidal-Madjar et al. 1994; Roberge et al. 
2002). \emph{IUE} observations of $\sigma$ Her have revealed the presence of 
narrow time-variable \ion{Si}{2}$^{*}$ $\lambda$ 1533.4 absorption features 
which are blueshifted with respect to the heliocentric velocity of the star 
(-11 km/sec; Bruhweiler, Grady, \& Chiu 1989) with a range of 2 to 21 km/sec 
(Grady et al. 1991). Although $\sigma$ Her possess time-variable, 
velocity-shifted absorption features similar to $\beta$ Pic, the high stellar 
luminosity and binary nature of the system may result in a different fate for 
gas liberated around $\sigma$ Her compared with material liberated around 
$\beta$ Pic.

\section{OBSERVATIONS}
Our data were obtained on 2001 July 15 using the \emph{Far Ultraviolet 
Spectroscopic Explorer} (\emph{FUSE}). Our observations were made in histogram 
mode, using the low resolution 30$\arcsec$ $\times$ 30$\arcsec$ aperture which 
covers the wavelength range between 905 and 1187 \AA. The \emph{FUSE} satellite
has four channels (LiF 1, SiC 1, LiF 2, SiC 2) which form two nearly identical 
``sides'' (labeled 1 and 2) and each consist of a LiF grating, a SiC grating 
and, a detector. Each detector is divided into two independent segments (a and 
b), seperated by a small gap. The eight partially overlapping spectra that fall
on different portions of the two detectors (LiF 1a, LiF 1b, etc.) cover the 
entire wavelength range. A description of the on orbit performance of 
\emph{FUSE} is described in (Sahnow et al. 2000) with a conservative estimate 
for the spectral resolution (R = $\lambda / \Delta \lambda$ = 15000) of the 
spectrograph. The data were calibrated at Johns Hopkins using the CALFUSE 2.0.5
pipeline. 

We wavelength calibrate our data by cross correlating the spectrum of $\sigma$
Her with a rotationally broadened synthetic stellar spectrum (Chayer 2001) 
in order to determine the relative wavelength of the circumstellar lines
with respect to the stellar lines. Since the synthetic spectrum (calculated for
a star with solar metallicty, $T_{eff}$ = 10,000 K, log $g$ = 4.0 and $v\sin i$
= 270 km/sec) fits the observed spectrum well in the region of the stellar 
\ion{C}{1} absorption lines (which lie in the LiF 1B channel), we are able
to set the ``zero-point'' wavelength calibration for the spectrum to within
15 km/sec. Since the LiF 1A and LiF 1B channels have the same wavelength shift,
cross correlating the observed LiF 1B channel spectrum with the synthetic 
spectrum allows us to calculate the ``zero-point'' shift needed for both LiF 1A
and LiF 1B. The offsets for the remaining channels analyzed (LiF 2B, SiC 1A, 
SiC 2B) are calculated by cross correlating these spectra with spectra from the
LiF 1A channel in the wavelength regions in which they overlapped. The 
uncertainty in the relative wavelength calibration is dominated by detector 
distortions and is $\sim$9 km/sec. No circumstellar lines were detected at 
wavelengths within the SiC 2A/1B channels (905 \AA to 1005 \AA); thus, these 
channels are not discussed further in this paper.

All of our LiF 2A channel exposures of $\sigma$ Her suffer from an effect 
called detector x-walk, in which low pulse height events are misplaced in the 
dispersion direction because of gain sag in the regions on the detector
on which the brightest airglow lines fall. Currently, there is no correction
for detector x-walk in the histogram data calibration pipeline; thus, we 
excluded all LiF 2A spectra from our analysis.

\section{CIRCUMSTELLAR GASES}
Absorption in the excited fine structure states, \ion{C}{2}$^{*}$ at 
1037.02 \AA, \ion{N}{2}$^{*}$ at 1084.58 \AA, and \ion{N}{2}$^{**}$ at 
1085.54 \AA \ and 1085.70 \AA \ are observed (see Figure 1a and 1b). Since 
these lines are significantly narrower than the rotationally broadened 
stellar lines and broader than the observed interstellar lines, shown for 
comparison in Figures 1c and 1a respectively, the gases are circumstellar. 
No circumstellar \ion{O}{1} is detected toward $\sigma$ Her.

We fit a model to each multiplet using a quadratic polynomial plus a Gaussian
to represent the photospheric component and Gaussians to represent each of the
narrow circumstellar components. We initially fit the model by hand, masking 
out the narrow circumstellar lines and fitting the broad photospheric 
component and adding each circumstellar component back individually. Then, we 
ran a minimize $\chi^{2}$ routine on the model using our initial fit as a 
starting point. In each model fit, we let the wavelength, scale, and FWHM for 
each Gaussian be free parameters in addition to the 3 coeffecients for the 
quadratic polynomial. The model fits for the \ion{C}{2} and \ion{N}{2} features
are overlaid on the spectra shown in Figures 1a and 1b and have $\chi^{2}$ = 
0.77 and 0.91 respectively. Changing the wavelength of the circumstellar
features by 10 km/sec and 20 km/sec increases the reduced $\chi^{2}$ of the fit
to $\sim$1.5 and $\sim$2.5 respectively for \ion{C}{2} and to $\sim$1.4 and
$\sim$2.4 respectively for \ion{N}{2}. We use the wavelengths from the fits 
of the narrow circumstellar cores to determine the velocity shift of the gas 
and the photosphere fit to estimate the continuum in our measurements of the 
equivalent widths. The excited \ion{N}{2}$^{**}$ appears blueshifted by 
25-28 km/sec and the excited \ion{C}{2}$^{*}$ appears blueshifted by 20 
km/sec. The line equivalent widths and their statistical uncertainities are 
given in Table 2. The population of all observed species in excited fine 
structure levels can be explained with radiative pumping by stellar 
ultraviolet photons. 

We observed no absorption in the H$_{2}$ Lyman (5,0) band or the CO C-X (0,0) 
band toward this star. The 3$\sigma$ upper limits on the H$_{2}$ line-of-sight 
column densities in the J = 0 and J = 1 rotational levels are measured, from 
the P(1) and R(0) transitions at 1036.54 \AA \ and 1038.16 \AA, to be 
$N(J = 0) \leq 4 \times 10^{14}$ cm$^{-2}$ and $N(J = 1) \leq 7 \times 10^{14}$
cm$^{-2}$ using wavelengths and oscillator strengths from Abgrall et al (1993).
The 3$\sigma$ upper limits on the CO line-of-sight column densities 
in the J = 2 and J = 4 rotational levels are measured, from the P(3) and
R(3) transitions at 1088.05 \AA \ and 1087.73 \AA, to be $N(J = 2) \leq 1.4 
\times 10^{14}$ cm$^{-2}$ and $N(J = 4) \leq 1.0 \times 10^{14}$ cm$^{-2}$
using wavelengths and oscillator strengths from Morton \& Noreau (1994).
The upper limits on the column density of CO in the J = 2 and J = 4 levels
suggest that $N(CO) \leq 5 \times 10^{14}$ cm$^{-2}$ or $N(CO) \leq 7 \times 
10^{14}$ cm$^{-2}$ for excitation temperatures of 25 K and 100 K respectively.

\section{PARENT BODIES}
The observation of mid infrared excess associated with $\sigma$ Her suggests
the presence of micron-sized grains around this star. We hypothesize that the 
circumstellar gas detected around $\sigma$ Her is physically associated with 
with the dust in ways described below.

Since $\sigma$ Her is a B9V star, radiation pressure will effectively
remove small grains from the environment around the star. A lower limit to the 
size of dust grains orbiting a star can be found by balancing the force due to 
radiation pressure with the force due to gravity. For small grains with radius
$a$, the force due to radiation pressure overcomes gravity for:
\begin{equation}
a < 3 L_{*} Q_{pr}/(16 \pi G M_{tot} c \rho_{s})
\end{equation}
(Artymowicz 1988) where $L_{*}$ is the stellar luminosity, $M_{tot}$ is the
binary mass, $Q_{pr}$ is the radiation pressure coupling coefficient, and 
$\rho_{s}$ is the density of an individual grain. Since radiation from a late 
B-type star is dominated by optical and ultraviolet light, we expect that 
$2 \pi a/\lambda \gg 1$ and therefore the effective cross section of the 
grains can be approximated by their geometric cross section so 
$Q_{pr} \approx 1$. We estimate the stellar luminosity from the bolometric 
magnitude using the \emph{Hipparcos} V-band magnitude ($m_{V}$ = 4.01 mag),
correcting for the interstellar extinction measured along the line of sight 
(E(B-V)=0.06; Fajardo-Acosta et al. 1998) using the far ultraviolet extinction 
law derived in Cardelli, Clayton, \& Mathis (1989), a distance 93 pc, and 
a bolometric correction (Flower 1996) corresponding to an effective temperature
$T_{eff}$ = 10,500 K. For the primary, we estimate a stellar luminosity $L_{*}$
= 230 $L_{\sun}$. We similarly estimate the luminosity for the secondary 
assuming a relative magnitude $\delta m$ = 2.5 mag and a bolometric correction 
corresponding to an effective temperature of $T_{eff}$ = 9530 K. For the
secondary, we find $L_{*}$ = 7.4 $L_{\sun}$. Thus, the contribution of the 
secondary to the total luminosity of the system is small and can be neglected. 
The mass of the binary system has been determined astrometrically (M$_{tot}$ = 
4.5 $\pm $0.8 M$_{\sun}$; Martin et al. 1998). For circumstellar dust grains 
around $\sigma$ Her with a density $\rho_{s}$ = 2.5 g cm$^{-3}$, we find a 
minimum grain radius of $a$ = 15 $\mu$m; thus, the grains are large enough to 
act as black bodies. Grains smaller than 15 $\mu$m will be effectively removed 
by radiation pressure on a timescale of $<$10 years.

Another mechanism which may remove particles from the circumstellar environment
is Poynting-Robertson drag. The Poynting-Robertson lifetime of grains in a 
circular orbit, a distance $D_{in}$ from a star is
\begin{equation}
t_{PR} = \left( \frac{4 \pi a \rho_{s}}{3} \right) \frac{c^{2} D_{in}^{2}}{L_{*}}
\end{equation}
(Burns et al. 1979). With the parameters given above and $D_{in}$ = 20 AU, the 
Poynting-Robertson lifetime of the grains is $t_{PR}$ = 4.6$\times 10^{4}$ 
years. Since this timescale is significantly shorter than the stellar age, 
$t_{age}$, we hypothesize that the grains are replenished through collisions 
between larger bodies. It is difficult to provide an accurate
estimate of the true mass in parent bodies. Since $\sigma$ Her has an age of
$\sim$140 Myr, objects with a radius larger than 4.5 cm will not have had 
enough time to spiral into the star under Poynting-Robertson drag. Large 
masses of circumstellar dust could exist around $\sigma$ Her and be difficult 
to detect at mid infrared wavelengths because large grains have relatively 
less surface area compared to their mass.

We can estimate a lower limit for the total mass contained in parent bodies 
around $\sigma$ Her assuming that the system is in a steady state. If $M_{PB}$ 
denotes the mass in parent bodies, then we may write
\begin{equation}
M_{PB} \geq \frac{4 L_{IR} t_{age}}{c^2}
\end{equation}
(Chen \& Jura 2001). Assuming that the binary system $\sigma$ Her has a 
fractional infrared luminosity $L_{IR}/L_{*}$ = 6.6$\times$10$^{-5}$
(Fajardo-Acosta et al. 1998) and an age $t_{age}$ = 140 Myr, we find a
minimum mass in parent bodies of $M_{PB}$ $\geq$ 0.20 $M_{\earth}$.
By analyzing the composition of the circumstellar gas, we can infer the
properties of the parent bodies. 

\section{THE WIND FROM THE CIRCUMSTELLAR MATTER}
The approximation that the gas is optically thin is probably not valid for 
the \ion{C}{2}. If the atoms were optically thin, then the ratio of their 
equivalent widths, $W_{\lambda}$(\ion{C}{2})/$W_{\lambda}$(\ion{C}{2}$^{*}$), 
would be 0.50, assuming the excitation temperature is $\gg$100 K. However, 
from our \emph{FUSE} spectra, we measure 
$W_{\lambda}$(\ion{C}{2})/$W_{\lambda}$(\ion{C}{2}$^{*}$) = 0.83. We estimate 
the optical depths of the atoms in the ground state and first excited state
from the ratio of the measured equivalent widths, assuming normalized Gaussian
line profiles.
\begin{equation}
\phi(\Delta \nu) = \frac{1}{\Delta \nu_{o} \sqrt \pi} 
                   \exp(\frac{- \Delta \nu^{2}}{\Delta \nu_{o}^{2}})
\end{equation}
where $\Delta \nu_{o}$ is the line width. The equivalent width of a line with
a Gaussian profile is 
\begin{equation}
\frac{W_{\lambda}}{\lambda} = \frac{2 \Delta \nu_{o} F(\tau_{o})}{\nu_{o}} 
\end{equation}
where
\begin{equation}
F(\tau_{o}) = \int_{0}^{\infty} \left[1 - \exp(-\tau_{o} e^{-x^{2}}) \right] dx
\end{equation} 
(Spitzer 1978) and $\tau_{o}$ is the optical depth at line center. We 
estimate $\tau_{o}$ for \ion{C}{2} and \ion{C}{2}$^{*}$, assuming that 
$\tau_{o}$(\ion{C}{2})/$\tau_{o}$(\ion{C}{2}$^{*}$) = 0.5 and using the
observational result that 
F($\tau_{o}$(\ion{C}{2}))/F($\tau_{o}$(\ion{C}{2}$^{*}$)) = 0.83.
For \ion{C}{2} and \ion{C}{2}$^{*}$, we estimate $\tau_{o}$ = 3.8 and 
$\tau_{o}$ = 7.6 respectively and $\Delta \nu_{o}$ = 2.0 $\times$ 10$^{11}$ Hz.
In Table 2, we estimate the column densities of \ion{C}{2} and 
\ion{N}{2} using the equivalent widths given in Table 2 and wavelengths 
and oscillator strengths from Morton (1991), assuming that the \ion{N}{2} is
optically thin and that the \ion{C}{2} has the optical depths given above
and is located at the edge of the cloud. It is difficult to determine the
optical depth of the \ion{N}{2} lines because the \ion{N}{2}$^{*}$ and 
\ion{N}{2}$^{**}$ lines, which are observed with \emph{FUSE}, are doublet
and triplet blends. High resolution spectroscopy is needed to determine
the detailed characteristics of the gas. 

Whether atomic gas is expected to fall into the star or to be blown out of the
circumstellar environment can be determined from $\beta$ = 
$F_{rad}$/$F_{grav}$. We calculate the ratio of the force due to radiation 
pressure to the force due to gravity for the observed atomic species assuming 
that the \ion{N}{2} is optically thin and the \ion{C}{2} has the optical
depth given above and is at the edge of the cloud. The force due to radiation 
pressure acting on the atoms is given by the following expression: 
\begin{equation}
F_{rad} = \sum_{all \ transitions} \frac{\pi e^2}{m_{e} c^2} f_{i} G_{i}(\tau)
	  \Phi_{o_{i}},
\end{equation}
where  $f_{i}$ is the absorption oscillator strength, $\Phi_{o_{i}}$ is 
the flux of photons arriving at the distance of the atom from the star,
and $G(\tau)$ is the line shielding function.
\begin{equation}
G(\tau) = \int_{-\infty}^{\infty} \phi(\Delta \nu) 
          \frac{\Phi_{\nu}(\Delta \nu)}{\Phi_{o}} d(\Delta \nu)
\end{equation}
where $\Phi_{\nu}(\Delta \nu)$ is the flux in the cloud at $\Delta \nu$.
\begin{equation}
\Phi_{\nu}(\Delta \nu) = \Phi_{o} \exp(-\tau(\Delta \nu))
\end{equation}
If the gas is optically thin at all frequencies, then G($\tau$) = 1.0. We use 
an approximation for $G(\tau)$ given by Federman, Glassgold, \& Kwan (1979). 
For $\tau$ = 3.8 and 7.6, $G(\tau)$ = 0.12 and $G(\tau)$ =  0.051 
respectively. We infer $\Phi_{o_{i}}$ from our \emph{FUSE} spectra and from 
archival \emph{IUE} spectra at longer wavelengths. Correcting for interstellar
extinction, as described in section 4, suggests that the stellar flux
is 2.3 and 2.1 times brighter than observed by \emph{FUSE} at the \ion{C}{2} 
and \ion{N}{2} line wavelengths respectively. The estimated values for $\beta$ 
are listed in Table 2. The radiation pressure for the ground state and excited 
\ion{N}{2} is generated by transitions in the \emph{FUSE} wavelength range 
while the radiation pressure for the ground state and excited \ion{C}{2} is 
dominated by high f-value transitions at $\sim$1300 \AA \ where the stellar 
flux is significantly stronger.  For \ion{C}{2} and \ion{N}{2} atoms around 
$\sigma$ Her, $\beta$ $>$ 1 suggesting that the high luminosity of the B9V 
primary effectively removes circumstellar gas from this system in a low 
velocity wind. 

We can calculate the outflow velocity for atomic gas around $\sigma$ Her 
assuming that radiation pressure and gravity act on the gas and making the 
simplification that the atoms begin from rest. Since radiation pressure and 
gravity are the only forces acting on the gas once it is released from the 
parent bodies, the gas is expected to be on a hyperbolic orbit. We derive 
an expression for the outflow velocity, $v_{\infty}$, from the equation for 
the force acting on the gas. If the gas is low density so that collisions are 
unimportant, then the outflow velocity, $v_{\infty}$, for $\beta$ $\geq$ 1, is 
\begin{equation}
v_{\infty}= \left[ \sum_{all \ transitions} 
            \frac{G_{i}(\tau) L_{\nu_{i}}e^{2} f_{i}}
                 {2 D_{in} m_{e} c^2 m_{atom}} 
            (1-\frac{1}{\beta}) \right] ^{\onehalf}
\end{equation}
where $m_{atom}$ is the mass of the atomic species. For all the observed 
states of \ion{N}{2}, the outflow velocity of atomic gas blown out of this 
system is $v_{\infty}$ = 24 - 26 km/sec, assuming $D_{in}$ = 20 AU and that 
the gas is optically thin. For \ion{C}{2} and \ion{C}{2}$^{*}$, the predicted 
outflow velocities are $v_{\infty}$ = 29 km/sec and $v_{\infty}$ = 31 km/sec 
respectively, assuming $D_{in}$ = 20 AU and that the gas is at the edge of the
cloud. These velocities are in rough agreement with the observed blueshifts of
\ion{C}{2}$^{*}$ and \ion{N}{2}$^{**}$. 

We can estimate the gas masses entrained in the wind around $\sigma$ 
Her based upon the measurements of the column densities and the escape 
velocities for the atomic species \ion{C}{2} and \ion{N}{2}. If the gas is 
distributed in a spherically symmetric shell about the star, the mass of 
gas carried away by the wind is
\begin{equation}
dM/dt \approx 4 \pi D_{in} m_{atom} N v_{\infty} 
\end{equation}
Summing over all observed states of \ion{C}{2} and \ion{N}{2} and assuming
the column densities and escape velocities in Table 2, we estimate
dM(\ion{N}{2})/dt = 3.5$\times$10$^{-12}$ $M_{\sun}$/yr and 
dM(\ion{C}{2})/dt = 2.8$\times$10$^{-12}$ $M_{\sun}$/yr if $D_{in}$ = 20 AU.
We can also estimate the \ion{N}{2} mass loss rate from the momentum imparted 
to the gas by radiation pressure, 
\begin{equation}
v_{\infty} \frac{dM(\textup{N II})}{dt} = \frac{L_{\lambda}}{c} W_{\lambda} 
\end{equation}
If $v_{\infty}$ = 28 km/sec, we estimate dM(\ion{N}{2})/dt = 1.5 $\times$ 
10$^{-12}$ $M_{\sun}$/yr, consistent with the estimate made assuming that the 
\ion{N}{2} is optically thin. 

If the system is in a steady state, the wind observed around $\sigma$ Her 
returns $\sim$ 100 M$_{\earth}$ of gas during the lifetime of the system. 
Thus, $\sigma$ Her depletes the mass in parent bodies in less than the 
estimated age of the star. Either the dust currently observed is produced in a
short-lived stochastic event, or the system is in a steady state and the true 
parent body mass is probably well above the minimum we derive. For example, 
this mass could exist as meter-sized objects which are in stable orbits about 
the binary system and are difficult to detect in the mid infrared. At 20 AU, 
the estimated mass loss rate implies an electron density, $n_{e}$ $\approx$ 
\.{M}/(4$\pi D_{in}^{2} v_{\infty} m_{atom}$) $\approx$ 8.0 cm$^{-3}$ when 
summed over the observed species, suggesting that electron-atom collisions are
only moderately significant in this environment. 

\section{DISCUSSION}
Although the circumstellar gas around $\sigma$ Her appears to be outflowing in
a wind, the physical processes which generate the gas are uncertain. One 
possibility is that the gas is generated from the sublimation of comets, as 
observed around $\beta$ Pic (Vidal-Madjar et al. 1994) and 51 Oph (Roberge et 
al. 2002). Another possibility is that the gas is produced through collisions 
between parent bodies. Here, we consider both models.

If orbiting comets produce the absorption features observed then there 
could be as many redshifted events as blueshifted events (Crawford, Beust, 
\& Lagrange 1998). However, all of the absorption features observed toward 
$\sigma$ Her are blueshifted. Futhermore, sublimation and subsequent 
photodissociation of water ice from comets should produce \ion{O}{1}. No 
\ion{O}{1} is observed in the \emph{FUSE} spectrum of $\sigma$ Her. Finally,
we can estimate the distance at which comets sublime around $\sigma$ Her
assuming typical gravitational motions.
\begin{equation}
D_{sub} \sim \frac{G M_{tot}}{(\Delta v)^2}
\end{equation}
where $\Delta v$ is the observed line width in velocity units. For $\sigma$ 
Her, $D_{sub}$ = 9 AU assuming $M_{tot}$ = 4.5 $M_{\sun}$ and $\Delta v$ = 21 
km/sec neglecting radiative effects. This distance is thirty times further 
away then distance at which comets sublime around $\beta$ Pic (Beust et al. 
1998). In order to produce the same column density of absorbers, there must be
1000 times more comets around $\sigma$ Her compared to $\beta$ Pic (Beust, 
Karmann, \& Lagrange 2001). The existence of so many comets around $\sigma$ 
Her seems unlikley because the fractional infrared luminosity of $\sigma$ Her 
is only 0.02 times that of $\beta$ Pic.

Collisions between parent bodies may explain our data. 3-body orbits around 
$\sigma$ Her are expected to become dynamically unstable at distances less 
than approximately three times the binary separation (Artymowicz \& Lubow 
1994). For $\sigma$ Her, with a binary separation of 7 AU, this corresponds 
to $\sim$20 AU. Here, unstable orbits begin to cross each other, producing 
collisions between different objects. These collisions could provide a natural
explanation for the generation of atomic gas from circumstellar matter. The 
high luminosity of the B9V primary then drives the gas from the circumstellar 
environment in a wind. 

\section{CONCLUSIONS}
We have obtained an ultraviolet spectrum (between 905 \AA \ and 1187 \AA) of 
the nearby binary system $\sigma$ Her using \emph{FUSE}. We argue the 
following:

1. The ultraviolet spectrum of $\sigma$ Her possess absorption in the excited 
fine structure lines of \ion{C}{2} and \ion{N}{2}. The excitation of these 
states and the narrow width of these absorption features suggest that the gas 
is circumstellar. 

2. Since the Poynting-Robertson drag lifetime of dust grains at $D_{in}$ =
20 AU with $a$ = 15 $\mu$m is $4.6\times10^{4}$ yr, significantly shorter than 
the estimated age of $\sigma$ Her, the grains must be replenished from a 
reservoir such as collisions between larger objects. For $\sigma$ Her, we 
estimate that the minimum mass of parent bodies is 0.20 $M_{\earth}$.

3. Because $\sigma$ Her is a binary with a separation of 7 AU, parent body 
orbits become unstable at distances $\sim$20 AU from the system. Collisions 
of parent bodies at this distance could liberate gas which is then blown out 
of this system by the high luminosity of $\sigma$ Her.

4. If the gas is released from grains at $\sim$20 AU from the star, then the 
predicted outflow velocities of \ion{C}{2}$^{*}$ and \ion{N}{2}$^{**}$ are
 $\sim$31 km/sec and $\sim$26 km/sec respectively, in rough agreement with the
observed blueshifts of $\sim$20 km/sec and $\sim$28 km/sec respectively. 

5. We infer a mass loss rate of dM/dt $\sim$ 6 $\times$ 10$^{-12}$ 
M$_{\sun}$/yr, suggesting that $\sigma$ Her depletes the mass in parent bodies
in less than the estimated age of the system. This raises the liklihood that 
objects larger than $\sim$1 m are in orbit in this system.

\acknowledgements

This work has been supported by funding from NASA. We thank P. Chayer for 
providing us with a rotationally broadened synthetic spectrum for $\sigma$
Her. We also thank A. Roberge for her comments and B.-G. Anderson for running 
our data through the CALFUSE 2.0.5 pipeline.

\begin{deluxetable}{lll}
\singlespace
\tablecaption{$\sigma$ Her Properties} 
\tablehead{
    \colhead{Quantity} &
    \colhead{Adopted Value} &
    \colhead{Reference} \\
}
\tablewidth{0pt}
\tablecolumns{3}
\startdata
    Primary Spectral Type & B9V & 1 \\
    Distance & 93 pc & 2 \\
    Effective Temperature (T$_{eff}$) & 10,500 K & \\
    Stellar Radius (R$_{*}$) & 4.6 R$_{\sun}$ & \\
    Stellar Luminosity (L$_{*}$) & 230 L$_{\sun}$ & \\
    Total Binary Mass (M$_{tot}$) & 4.5 M$_{\sun}$ & 3 \\
    Rotational Velocity ($v\sin i$) & 270 km/sec & 1 \\
    Fractional Dust Luminosity & 6.6$\times$10$^{-5}$ & 4 \\
        \ \ \ \ \ ($L_{IR}/L_{*}$) & & \\ 
    Estimated Age & 140 Myr & \\
    Inner Grain Temperature & 240 K & 4 \\
    Inner Dust Distance ($D_{in}$) & 20 AU & \\
    Outer Grain Temperature & 100 K & \\
    Outer Dust Distance ($D_{out}$) & 120 AU & \\
    Minimum Parent Body Mass ($M_{PB}$) & 0.20 $M_{\earth}$ & \\
\enddata
\tablerefs{(1) Hoffleit \& Warren (1991);
           (2) \emph{Hipparcos};
           (3) Martin et al. (1998);
           (4) Fajardo-Acosta et al. (1998)
          }
\end{deluxetable}

\begin{deluxetable}{lccccccc}
\singlespace
\tablecaption{Properties of Circumstellar Gas} 
\tablehead{
    \omit &
    \omit &
    \omit &
    \omit &
    \omit &
    \colhead{Observed} &
    \colhead{Predicted} &
    \omit \\
    \colhead{Species} &
    \colhead{Wavelength} & 
    \colhead{Ground State Energy} & 
    \colhead{$W_{\lambda}$\tablenotemark{\dagger}} & 
    \colhead{N\tablenotemark{\ddagger}} & 
    \colhead{$v_{\infty}$} &
    \colhead{$v_{\infty}$\tablenotemark{\ddagger}} &
    \colhead{$\beta$}\\
    \omit & 
    \colhead{(\AA)} & 
    \colhead{(cm$^{-1}$)} & 
    \colhead{(m\AA)} & 
    \colhead{(10$^{14}$ cm$^{-2}$)} & 
    \colhead{(km/sec)} &
    \colhead{(km/sec)} &
    \omit \\ 
}
\tablewidth{0pt}
\tablecolumns{7}
\startdata
    \ion{C}{2} & 1036.34 & 0.00 & 184 $\pm$ 12 & 4.1 & -16 & -29 & 3.2 \\
    \ion{C}{2}$^{*}$ & 1037.02 & 63.42 & 222 $\pm$ 14 & 8.2 & -20 & -31 & 3.4 \\
    \ion{N}{2} & 1083.99 & 0.00 & 109 $\pm$ 10 & $\geq$ 1.0 & -14 & -24 & 2.5 \\
    \ion{N}{2}$^{*}$ & 1084.58 & 48.67 & 158 $\pm$ 17 & $\geq$ 1.5 & -28 & -24 & 2.5 \\
    \ion{N}{2}$^{**}$ & 1085.54 & 130.8 & 114 $\pm$ 12 & $\geq$ 6.7 & -25 & -26 & 2.7 \\
    \ion{N}{2}$^{**}$ & 1085.70 & 130.8 & 221 $\pm$ 21 & $\geq$ 2.5 & -28 & -26 & 2.7 \\ 
    \ion{O}{1}$^{*}$& 1040.94 & 158.27 & $\leq$ 39 & $\leq$ 4.5 & N/A & & \\
    \ion{O}{1}$^{**}$ & 1041.69 & 226.98 & $\leq$ 31 &  $\leq$ 3.5 & N/A & & \\
    H$_{2}$ (J=0) & 1036.54 & 0.00 & $\leq$ 95 & $\leq$ 4 & N/A & & \\
    H$_{2}$ (J=1) & 1038.16 & 118.16& $\leq$ 54 & $\leq$ 7 & N/A & & \\
    CO (J=2) & 1088.05 & 11.58 & $\leq$ 75 & $\leq$ 1 & N/A & & \\
    CO (J=4) & 1087.73 & 38.61 & $\leq$ 70 & $\leq$ 1 & N/A & & \\
\enddata
\tablenotetext{\dagger}{error bars are statistical uncertainties in the 
  measurement of the equivalent width and do not account for uncertainty
  in continuum fitting}
\tablenotetext{\ddagger}{\ion{C}{2} column densities, predicted outflow 
  velocities, and $\beta$s are corrected for optical depth assuming that the 
  gas is at the edge of the cloud. \ion{N}{2} is assumed to be optically thin.}
\end{deluxetable}

\begin{figure}[ht]
\figurenum{1a}
\epsscale{1.0}
\plotone{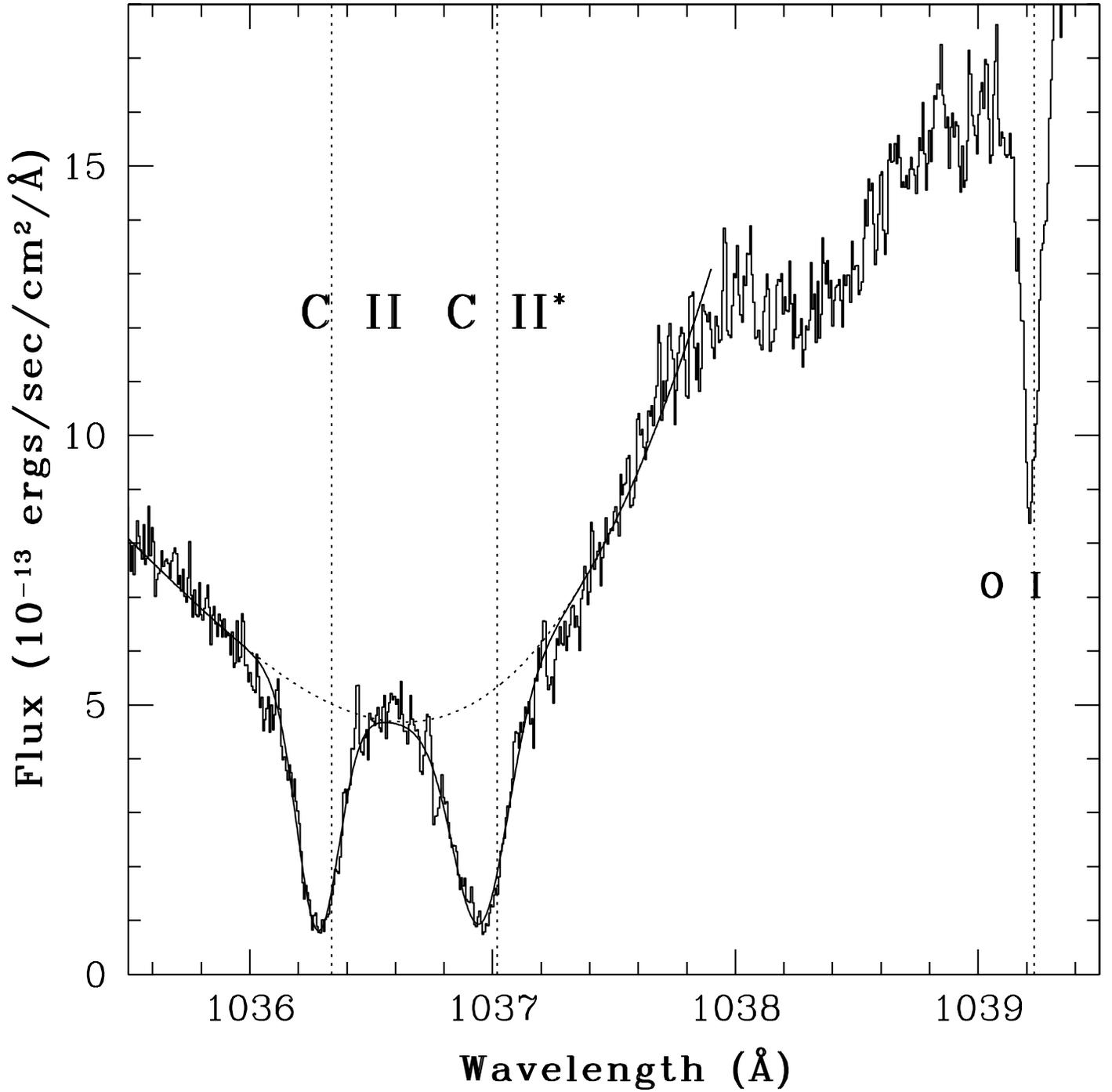}
\caption{The circumstellar \ion{C}{2} $\lambda$ 1037 doublet and interstellar
\ion{O}{1} $\lambda$ 1039 line. The solid curve is a minimum $\chi^{2}$ fit to 
the \ion{C}{2} doublet using Gaussians to model the photospheric and 
circumstellar components. The dotted curve shows the photosphere model. 
The vertical lines indicate wavelengths of the \ion{C}{2}, \ion{C}{2}$^{*}$,
and \ion{O}{1} lines at the star's velocity.}
\end{figure}

\begin{figure}[ht]
\figurenum{1b}
\epsscale{1.0}
\plotone{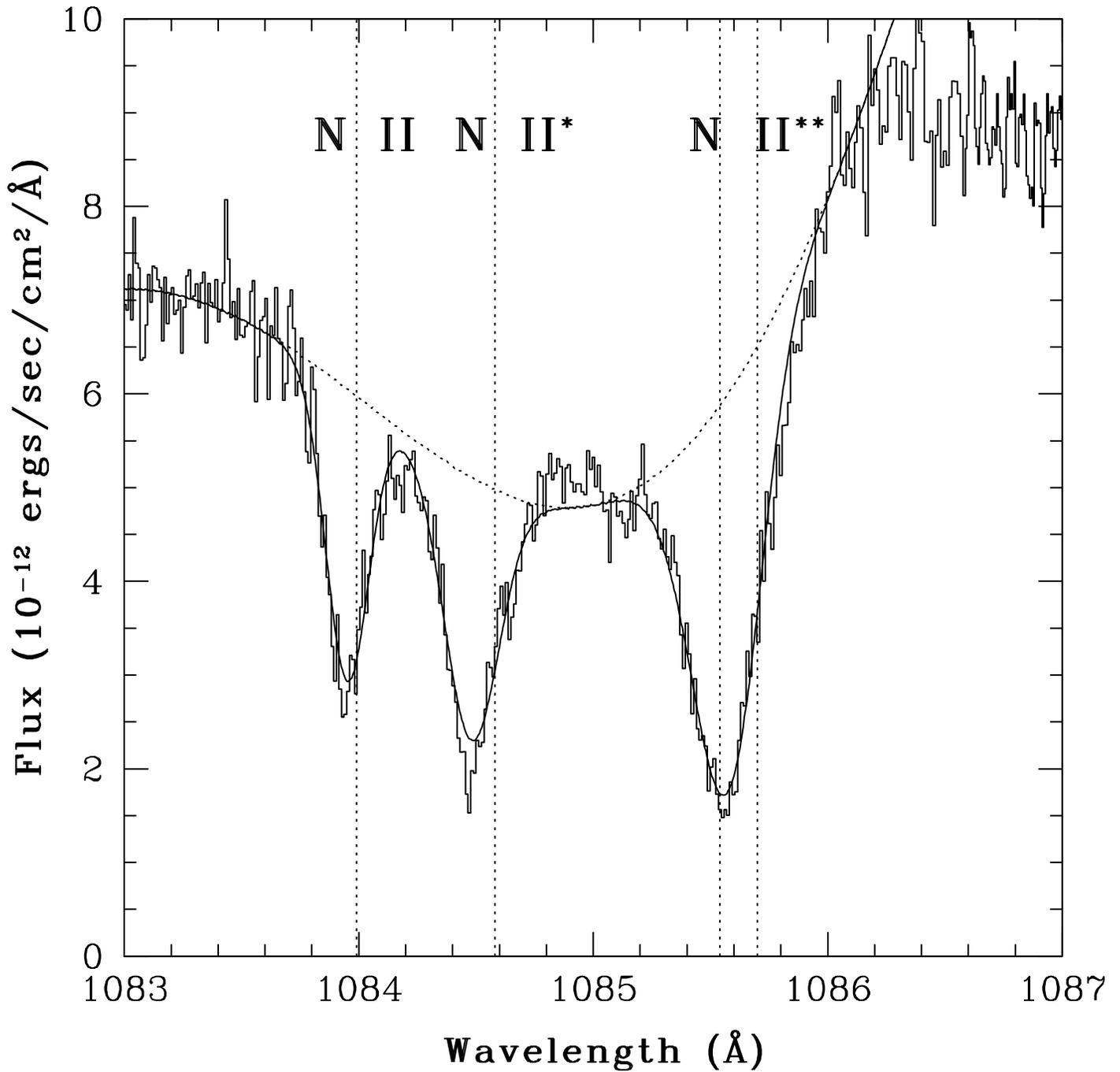}
\caption{The circumstellar \ion{N}{2} $\lambda$ 1085 multiplet. The solid 
curve is a minimum $\chi^{2}$ fit to the \ion{N}{2} multiplet using Gaussians
to model the photospheric and circumstellar components. The dotted curve shows 
the photosphere model. The \ion{N}{2}$^{*}$ and \ion{N}{2}$^{**}$ lines are
a doublet and triplet respectively whose components are blended together.
The vertical lines indicate the wavelength of the \ion{N}{2} line, the average
wavelength of the \ion{N}{2}$^{*}$ doublet, and the wavelengths of the 
\ion{N}{2}$^{**}$ lines with the largest f-values at the star's velocity.}
\end{figure}

\begin{figure}[ht]
\figurenum{1c}
\epsscale{1.0}
\plotone{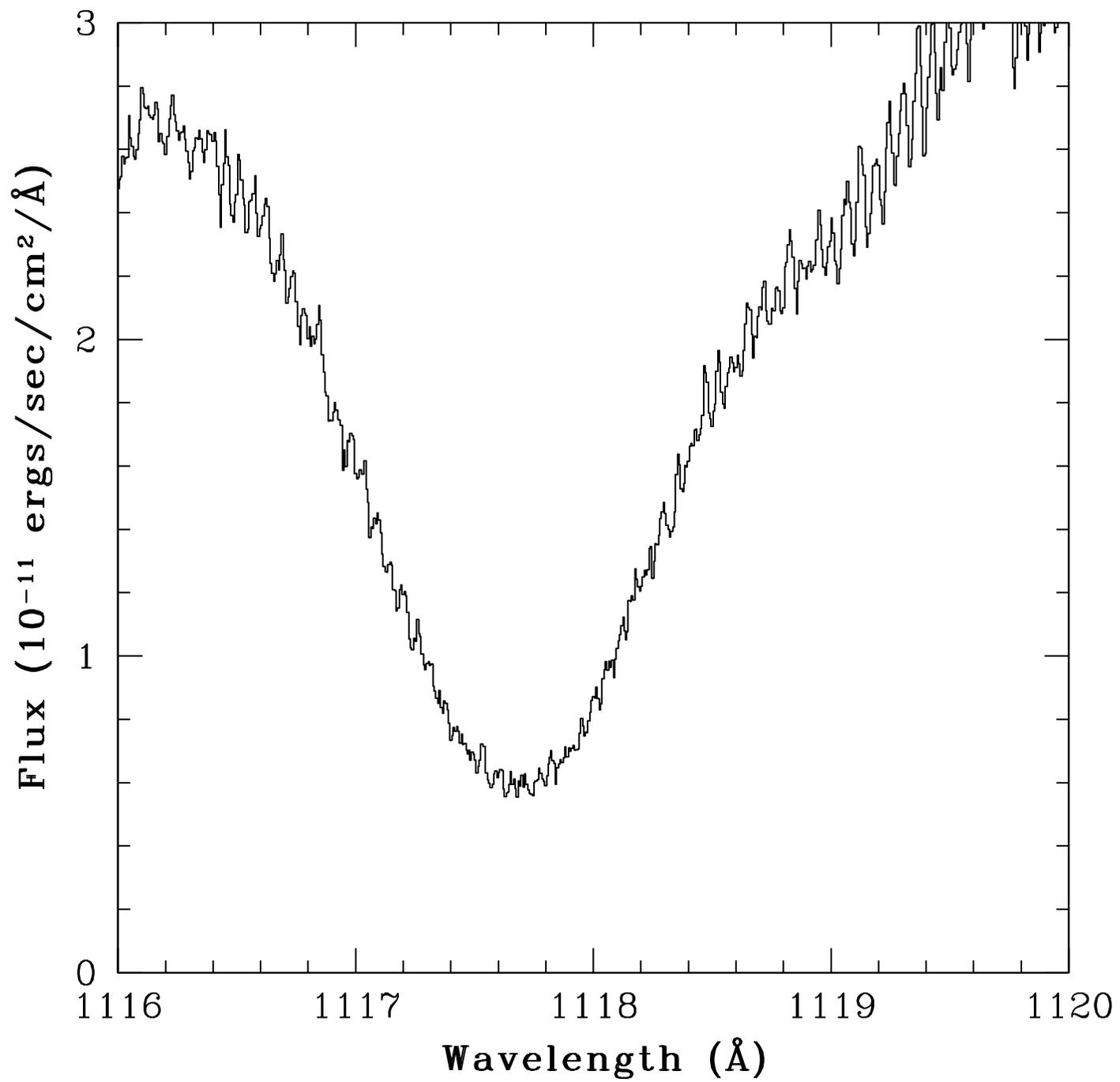}
\caption{The stellar \ion{C}{1} $\lambda$ 1118 line.}
\end{figure}

\end{document}